\documentclass[11pt,a4paper]{article}
\def\bea{\begin{eqnarray}}
\def\eea{\end{eqnarray}}
\def\nn{\nonumber}
\def\beq{\begin{equation}}
\def\eeq{\end{equation}}
\def\ba{\beq\new\begin{array}{c}}
\def\ea{\end{array}\eeq}
\def\be{\ba}
\def\ee{\ea}

\makeatletter
\newdimen\normalarrayskip              
\newdimen\minarrayskip                 
\normalarrayskip\baselineskip
\minarrayskip\jot
\newif\ifold             \oldtrue            \def\new{\oldfalse}
\def\arraymode{\ifold\relax\else\displaystyle\fi} 
\def\eqnumphantom{\phantom{(\theequation)}}     
\def\@arrayskip{\ifold\baselineskip\z@\lineskip\z@
     \else
     \baselineskip\minarrayskip\lineskip2\minarrayskip\fi}
\def\@arrayclassz{\ifcase \@lastchclass \@acolampacol \or
\@ampacol \or \or \or \@addamp \or
   \@acolampacol \or \@firstampfalse \@acol \fi
\edef\@preamble{\@preamble
  \ifcase \@chnum
     \hfil$\relax\arraymode\@sharp$\hfil
     \or $\relax\arraymode\@sharp$\hfil
     \or \hfil$\relax\arraymode\@sharp$\fi}}
\def\@array[#1]#2{\setbox\@arstrutbox=\hbox{\vrule
     height\arraystretch \ht\strutbox
     depth\arraystretch \dp\strutbox
     width\z@}\@mkpream{#2}\edef\@preamble{\halign
\noexpand\@halignto
\bgroup \tabskip\z@ \@arstrut \@preamble \tabskip\z@ \cr}%
\let\@startpbox\@@startpbox \let\@endpbox\@@endpbox
  \if #1t\vtop \else \if#1b\vbox \else \vcenter \fi\fi
  \bgroup \let\par\relax
  \let\@sharp##\let\protect\relax
  \@arrayskip\@preamble}
\def\eqnarray{\stepcounter{equation}%
              \let\@currentlabel=\theequation
              \global\@eqnswtrue
              \global\@eqcnt\z@
              \tabskip\@centering
              \let\\=\@eqncr
              $$%
 \halign to \displaywidth\bgroup
    \eqnumphantom\@eqnsel\hskip\@centering
    $\displaystyle \tabskip\z@ {##}$%
    \global\@eqcnt\@ne \hskip 2\arraycolsep
         $\displaystyle\arraymode{##}$\hfil
    \global\@eqcnt\tw@ \hskip 2\arraycolsep
         $\displaystyle\tabskip\z@{##}$\hfil
         \tabskip\@centering
    &{##}\tabskip\z@\cr}
\begingroup\ifx\undefined\newsymbol \else\def\input#1 {\endgroup}\fi
\input amssym.def \relax
\input amssym
\newfont{\hr}{msbm10}
\newfont{\ams}{msam10}

\font\numbers=cmss12
\font\upright=cmu10 scaled\magstep1
\def\stroke{\vrule height8pt width0.4pt depth-0.1pt}
\def\topfleck{\vrule height8pt width0.5pt depth-5.9pt}
\def\botfleck{\vrule height2pt width0.5pt depth0.1pt}
\def\Zmath{\vcenter{\hbox{\numbers\rlap{\rlap{Z}\kern 0.8pt\topfleck}\kern
2.2pt
                   \rlap Z\kern 6pt\botfleck\kern 1pt}}}
\def\Qmath{\vcenter{\hbox{\upright\rlap{\rlap{Q}\kern
                   3.8pt\stroke}\phantom{Q}}}}
\def\Nmath{\vcenter{\hbox{\upright\rlap{I}\kern 1.7pt N}}}
\def\Cmath{\vcenter{\hbox{\upright\rlap{\rlap{C}\kern
                   3.8pt\stroke}\phantom{C}}}}
\def\Rmath{\vcenter{\hbox{\upright\rlap{I}\kern 1.7pt R}}}
\def\Z{\ifmmode\Zmath\else$\Zmath$\fi}
\def\Q{\ifmmode\Qmath\else$\Qmath$\fi}
\def\N{\ifmmode\Nmath\else$\Nmath$\fi}
\def\C{\ifmmode\Cmath\else$\Cmath$\fi}
\def\R{\ifmmode\Rmath\else$\Rmath$\fi}

\newcounter{app}

\def\app{\setcounter{equation}{0}
\def\theequation{\Alph{app}.\arabic{equation}}\par
   \addvspace{4ex}
   \@afterindentfalse
  \secdef\@app\@dapp}

\newcommand\@app{\@startsection {app}{1}{0ex}%
                                   {-3.5ex \@plus -1ex \@minus -.2ex}%
                                   {2.3ex \@plus.2ex}%
                                   {\normalfont\Large\bf}}
\def\@dapp#1{%
{\parindent \z@ \raggedright  \bf #1}\par\nobreak}
\def\l@app#1#2{\ifnum \c@tocdepth >\z@
    \addpenalty\@secpenalty
    \addvspace{1.0em \@plus\p@}%
    \setlength\@tempdima{8em}%
    \begingroup
      \parindent \z@ \rightskip \@pnumwidth
      \parfillskip -\@pnumwidth
      \leavevmode \bfseries
      \advance\leftskip\@tempdima
      \hskip -\leftskip
      #1\nobreak\hfil \nobreak\hb@xt@\@pnumwidth{\hss #2}\par
    \endgroup\fi}
\newcounter{sapp}[app]

\def\sapp{\def\theequation{\Alph{app}.\arabic{equation}}
\par
\@afterindentfalse
  \secdef\@sapp\@dsapp}
\newcommand{\@sapp}{\@startsection{sapp}{2}{\z@}%
                                     {-3.25ex\@plus -1ex \@minus -.2ex}%
                                     {1.5ex \@plus .2ex}%
                                     {\normalfont\large\bfseries}}

\def\@dsapp#1{%
{\parindent \z@ \raggedright  \bf #1
}\par\nobreak}
\newcommand{\l@sapp}{\@dottedtocline{2}{1.5em}{2.3em}}

\def\2{{1\over 2}}
\def\N2{${\cal N}=2$}

\def\be{ \begin{eqnarray} }
\def\ee{ \end{eqnarray} }

\def\bea{\begin{eqnarray}}
\def\eea{\end{eqnarray}}
\def\nn{\nonumber}

\def\beq{\begin{equation}}
\def\eeq{\end{equation}}
\def\ba{\beq\new\begin{array}{c}}
\def\ea{\end{array}\eeq}
\def\be{\ba}
\def\ee{\ea}

\def\arcsinh{\hbox{ arcsinh}}
\def\arccosh{\hbox{ arccosh}}

\title{p,q-Duality and Hamiltonian Flows
in the Space of Integrable Systems
\\
or
\\
Integrable Systems as Canonical Transforms
of the Free Ones
}

\author{A.Mironov$^{1,2}$\thanks{mironov@lpi.ac.ru, mironov@itep.ru},
A.Morozov$^{2}$\thanks{morozov@vx.itep.ru} \
\\ \normalsize \em $^{1}$ Theory
Department, Lebedev Physics Institute, Moscow
~117924, Russia
\\
\normalsize \em $^{2}$ITEP, Moscow
117259, Russia}

\date{}

\begin{document}

\maketitle

\vspace{-8.2cm}

\begin{center}
\hfill FIAN/TD-12/01\\
\hfill ITEP/TH-36/01\\
\hfill hep-th/0107114
\end{center}

\vspace{5.5cm}

\begin{abstract}

Variation of coupling constants of integrable system
can be considered as canonical transformation or,
infinitesimally,
a Hamiltonian flow in the space of such systems.
Any function $T(\vec p, \vec q)$ generates
a one-parametric family of integrable systems
in vicinity of a single system:
this gives an idea of how many integrable systems
there are in the space of coupling constants.
Inverse flow is generated by a dual ``Hamiltonian",
$\widetilde T(\vec p, \vec q)$
associated with the dual integrable system.
In vicinity of a self-dual point
the duality transformation just interchanges momenta and
coordinates in such a ``Hamiltonian":
$\widetilde T(\vec p, \vec q) = T(\vec q, \vec p)$.
For integrable system with several coupling constants the
corresponding ``Hamiltonians" $T_i(\vec p, \vec q)$
satisfy Whitham equations and
after quantization (of the original system) become
operators satisfying the zero-curvature condition in
the space of coupling constants:
$$
\left[\frac{\partial}{\partial g_a} - \hat T_a(\hat{\vec p},\hat{\vec q}),\
\frac{\partial}{\partial g_b} - \hat T_b(\hat{\vec p},\hat{\vec q})\right] = 0
$$
Some explicit formulas are given for harmonic oscillator and
for Calogero-Ruijsenaars-Dell system.

\end{abstract}


\section{Introduction}

The study of non-perturbative quantum phenomena revealed the
real role of integrable systems in physics:
$\tau$-functions of integrable hierarchies appear to
describe partition functions of quantum theories,
with time-variables identified with coupling constants
\cite{tauparrev, tauparexa}.
Broadening understanding of this relation, together with the
growing interest to peculiarities of essential (non-perturbative)
quantum physics, stimulates a new attention to the basics of
integrability theory, which brings one back to the very foundations
of {\it classical} Hamiltonian mechanics that were not sufficiently
investigated in the past.
The present paper is devoted to one of such basic subjects, to the
description of integrable systems in terms of the Hamiltonian evolution
in the space of coupling constants. This approach can help one to
describe the entire variety of integrable systems and to reveal the
meaning of peculiar relations, like dualities, between different such
systems.

\section{Hamiltonian flow in the space of integrable systems \label{Haf}}

\subsection{Variation of couplings as canonical transform -- peculiar feature
of integrable systems}

Integrable system with $N$ coordinates $q_i$ and $N$ momenta $p_i$
is characterized by existence of $N$ Poisson-commuting Hamiltonians
$H_i(\vec p,\vec q)$,
$\{H_i, H_j\} = \frac{\partial H_i}{\partial \vec p}\frac{\partial H_j}{\partial \vec q} -
\frac{\partial H_j}{\partial \vec p}\frac{\partial H_i}{\partial \vec q} = 0$.
For such a system one can consider a canonical transformation,
treating these Hamiltonians as new momenta-like or coordinate-like variables.
This transformation (of which the infinitesimal version
is a certain Hamiltonian flow) will be the subject of the present paper.

To make the problem precise, let us consider a one-parametric family of
integrable models, parameterized by a single coupling constant $g$ such that
the model is {\it free} when $g=0$.
This means that at $g=0$ the Hamiltonians
$H^{(0)}_i(\vec p) = H_i(\vec p,\vec q| g=0)$ are functions only of
momenta $\vec p$, though, for conventional choices of Hamiltonians in
particular applications, these functions can be non-trivial.
The typical examples are:
$H^{(0)}_k(\vec p) = \sum_{i=0}^N p_i^k$ and
$H^{(0)}_k(\vec p) = \sum_{I: |I|=k} \prod_{i \in I} e^{p_i}$
for $p$-rational and $p$-trigonometric models respectively and some elliptic functions
of $\vec p$ for their elliptic generalizations.

The adequate definition of the new canonical variables
$\vec P_g = \vec P (\vec p,\vec q|g)$ and
$\vec Q_g = \vec Q (\vec p,\vec q|g)$\footnote{
In what follows we often omit the label $g$, implying that the
capital letters $P$ and $Q$ always denote the dressed
momenta and coordinates $P_g$ and $Q_g$.
}
is \cite{dual}

\be
H_i^{(0)}(\vec P) = H_i(\vec p,\vec q|g) \\
\widetilde H_i^{(0)}(\vec q) = \widetilde H_i(\vec Q,\vec P|g)
\label{newP}
\ee
where $\widetilde H_i(\vec p,\vec q|g)$ define the Hamiltonians
of the {\it dual} integrable system (for other discussions of this duality
see \cite{duality} and \cite{dual2}).
Note that in (\ref{newP}) they depend on the
``dressed" variables $\vec P$ and $\vec Q$ and, moreover, $\vec P$ and
$\vec Q$ are interchanged.
The shape of the dual Hamiltonians is dictated by the requirement
that the new variables $\vec P$ and $\vec Q$ are canonical, i.e.

\be
\sum_i dP_i\wedge dQ_i = \sum_i dp_i\wedge dq_i
\ee
and the Poisson brackets are

\be
\{\ ,\ \} = \sum_i \left(\frac{\partial}{\partial p_i}\otimes
\frac{\partial}{\partial q_i} -
\frac{\partial}{\partial q_i}\otimes \frac{\partial}{\partial p_i}\right) =
\sum_i \left(\frac{\partial}{\partial P_i}\otimes
\frac{\partial}{\partial Q_i} -
\frac{\partial}{\partial Q_i}\otimes \frac{\partial}{\partial P_i}\right)
\ee

\subsection{Infinitesimal version of canonical transforms --
the Hamiltonian flows}

Relations (\ref{newP}) define $\vec P_g$ as functions of $\vec p,\vec q$ and
the coupling constant $g$.
Of special interest and importance is the infinitesimal version of this
{\it canonical} transformation, considered as a Hamiltonian flow along the
$g$ direction in the space of coupling constants.
Such transformation is generated by a new Hamiltonian $T(\vec p,\vec q; g)$,
according to the rule:

\be
\frac{\partial \vec P_g}{\partial g} =
\left\{ T(\vec P_g, \vec Q_g| g), \vec P_g \right\} = -\frac{\partial T}{\partial \vec Q_g}, \\
\frac{\partial \vec Q_g}{\partial g} =
\left\{ T(\vec P_g, \vec Q_g| g), \vec Q_g \right\}  = \frac{\partial T}{\partial \vec P_g}
\ee
This new Hamiltonian does not commute with the old ones, but it converts them
into a new set of commuting Hamiltonians.

Moreover, instead of considering a pre-given family of integrable systems,
one can use {\it any} function
$T(\vec p,\vec q| g)$ to define a whole one-parametric family of
integrable systems, though explicit construction of the corresponding
Poisson-commuting Hamiltonians is rarely possible.
Further, a multi-parametric family of integrable systems is similarly
generated by {\it any} collection of functions $T_a(\vec p,\vec q| g)$
satisfying the compatibility (Whitham) equations:\footnote{
As mentioned in the abstract to this paper,
after quantization of the original system these $T_a(\vec p,\vec q|\{g_b\})$
become a family of operators  $\hat T_a(\hat{\vec p},\hat{\vec q}|\{g_b\})$,
satisfying the zero-curvature condition in the space of coupling constants:
$$
\left[\frac{\partial}{\partial g_a} - \hat T_a(\hat{\vec p},\hat{\vec q}),\
\frac{\partial}{\partial g_b} - \hat T_b(\hat{\vec p},\hat{\vec q})\right] = 0
$$
This can serve as an explanation of emergency of {\it classical} integrability
(the zero-curvature equations) in the study of quantum integrable systems.
}

\be
\frac{\partial T_b}{\partial g_a} - \frac{\partial T_a}{\partial g_b} +
\left\{ T_a, T_b \right\} = 0
\label{PC}
\ee

The coupling constant variation for the dual integrable system is governed by
the dual Hamiltonian $\widetilde T(\vec p,\vec q|g)$: for

\be
\widetilde H_i^{(0)}(\widetilde{\vec P}) = \widetilde H_i(\vec p,\vec q|g)
\ee
(and
$H_i^{(0)}(\vec q) = H_i(\widetilde{\vec Q},\widetilde{\vec P} |g)$)
we have

\be
\frac{\partial \widetilde{\vec P_g}}{\partial g} =
\left\{ \widetilde T(\widetilde{\vec P_g}, \widetilde{\vec Q_g}| g), \widetilde{\vec P_g} \right\} =
-\frac{\partial \widetilde T}{\partial \widetilde{\vec Q_g}}, \\
\frac{\partial \widetilde{\vec Q_g}}{\partial g} =
\left\{ \widetilde T(\widetilde{\vec P_g}, \widetilde{\vec Q_g}| g), \widetilde{\vec Q_g} \right\}  =
\frac{\partial \widetilde T}{\partial \widetilde{\vec P_g}}
\ee

Relation between $\widetilde T(\vec p,\vec q|g)$ and $T(\vec p,\vec q|g)$
becomes especially simple at any self-dual point
(where $\widetilde H_i(\vec p,\vec q\ | g_{SD}) = H_i(\vec p,\vec q\ | g_{SD})$):

\be
\widetilde T(\vec p,\vec q\ | g_{SD}) = T(\vec q,\vec p\ | g_{SD})
\ee
As immediate consequence, {\it any} symmetric function
$T(\vec p,\vec q|g) = T(\vec q,\vec p|g)$ defines a one-parametric family
of self-dual integrable systems.

Note that it is not that immediate calculation to restore the Hamiltonian
starting from arbitrarily given $T(\vec p,\vec q|g)$. Say, one can build a
perturbative series in the coupling constant $g$. At the self-dual point
for the system with one degree of freedom one takes the expansion $
T(p,q|g)\equiv \sum_{i=1} T_i(p,q)g^{i-1}$,
$P=H(p,q)=p+\sum_{i=1}\Phi_i(p,q)g^i$ etc and obtains that
$\Phi_1(p,q)=\partial_q T_1(p,q)$, $\Phi_2(p,q)=\partial_q T_2(p,q)+...$.
The result ultimately is that the symmetric part of
$\partial_p\Phi_i(p,q)=\partial^2_{q,p} T_i(p,q)+$ is given independently at any order,
while the antisymmetric parts are fixed by $\Phi_i$'s at lower orders:

\be\label{sdg}
\Phi_1'(p,q)-\Phi_1'(q,p)=0,\\
\Phi_2'(p,q)-\Phi_2'(q,p)=\Phi_1''(p,q)\Phi_1(p,q)-(p\leftrightarrow q),\\
\Phi_3'(p,q)-\Phi_3'(q,p)=\Phi_1''(p,q)\Phi_2(p,q)+\Phi_2''(p,q)\Phi_1(p,q)-
{1\over 2}\Phi_1'''(p,q)\Phi_1^2(p,q)-\\-\Phi_1''(p,q)\Phi_1'(p,q)\Phi_1(p,q)
-(p\leftrightarrow q),\\
\ldots
\ee
where all the derivatives are taken w.r.t. to the first variable. A trivial
solution to these equations (\ref{sdg}) is $\Phi_i(p,q)=p^{n_i+1}q^{n_i}$
with arbitrary $\{n_i\}$. This implies that the Hamiltonian $H(p,q)=pf(pq)$
is self-dual with arbitrary function $f$. Indeed, one easily checks this is
the case (note that for such a Hamiltonian $pq=PQ$). We discuss later in
detail the rational Calogero Hamiltonian that gets to this class of
Hamiltonians.

\subsection{Generating functions of canonical transforms}

Along with the Hamiltonians $T(\vec P, \vec Q | g)$ one can also consider
the generating functions of canonical transformations in question, like
$S(\vec Q, \vec q|g)$ or its Legendre transform
$F(\vec P, \vec q|g)=\vec P\vec Q - S(\vec Q,\vec q|g)$,
such that

\be
-\vec P = \frac{\partial S}{\partial \vec Q}, \ \
\vec p = \frac{\partial S}{\partial \vec q}
\label{Pp}
\ee
and

\be
T = \frac{\partial S}{\partial g}
\label{PC2}
\ee
Of course, for canonical transformation

\be
\frac{\partial P_i}{\partial q_j} + \frac{\partial p_j}{\partial Q_i} = 0,
\label{symder}
\ee
as implied by (\ref{Pp}),
but one should be more careful when drawing similar conclusion from
(\ref{PC2}): the second $g$ derivatives satisfy eq.(\ref{PC}),
because $g$-derivative
is taken at constant $\vec P_g$ and $\vec Q_g$, which themselves depend on $g$.

Similarly,

\be
-\vec Q = \frac{\partial F}{\partial \vec P}, \ \
\vec p = \frac{\partial F}{\partial \vec q}
\ee
At self-dual points $F(\vec P,\vec q | g_{SD}) = F(\vec q,\vec P |
g_{SD})$ is a symmetric function.

Note that

\be
{\partial F\over\partial g}=T
\ee
(with $p$ and $Q$ constant), i.e. $T$ is, in a sense, a more invariant
quantity than $S$ and $F$, which does not depend on the choice of
independent variables.

Another way to see it is to consider a flow from an integrable system with
coupling constant $g_1$ to the same system with coupling constant $g_2$.
Then, $T$ depends only on $g_2$, but not on $g_1$, while the generating
functions depend on both $g_1$ and $g_2$.

\subsection{Generating functions, evolution operator and eigenfunctions}

Basically, if reexpressed in terms of $\vec Q$ and $\vec q$ (instead of $\vec P$
and $\vec Q$) the Hamiltonian $T = \partial S/\partial g$. Therefore, $e^{iS}$ can be
considered as a kind
of an evolution operator (kernel) in the space of coupling constants,
which performs a canonical transformation from the
free system to the integrable one.
After quantization it can be symbolically represented as

\be
e^{i\hat S}(\vec Q,\vec q|g) = \oplus_{\lambda} |\psi^{(0)}_\lambda(\vec Q)\rangle
c_{\vec \lambda} \langle \psi_\lambda(\vec q)|
\label{evo}
\ee
where $|\psi_\lambda\rangle$ and $|\psi^{(0)}_\lambda\rangle$
are eigenfunctions of the system and $c_{\vec \lambda}$ are some
coefficients, depending on the spectral parameter $\vec \lambda$.
The evolution operator satisfies the quantum version of eq.(\ref{newP}),

\be
H^{(0)}(\partial/\partial \vec Q) =
e^{-i\hat S(\vec Q,\vec q|g)}
H(\partial/\partial \vec q,\vec q|g) e^{\hat iS(\vec Q,\vec q|g)}
\ee
Since eigenfunctions $|\psi^{(0)}_\lambda(Q)\rangle$ of a free system,
are just exponents of the spectral parameter $\vec\lambda$,
the dressed eigenfunctions $|\psi_\lambda(q)\rangle$ are basically
Fourier transforms of the evolution operator $e^{iS(Q,q)}$:

\be
\psi_\lambda(q) \sim \int e^{i\hat S(Q,q)} e^{i\lambda Q} dQ
\label{efeS}
\ee
Similarly,

\be
\psi_\lambda(q) \sim e^{i\hat F(P,q)} \delta(\lambda - P) dP \sim e^{i\hat
F(\lambda,q)}
\label{efeF}
\ee
Note that there is a freedom in solutions of eq.(2) to shift $\{Q_i\}$ by an
arbitrary function of $\{P_i\}$. This shift is quite complicated in terms of
the generating function $S$, but in $F$ it is just an addition of the term
depending only on $\{P_i\}$. In the quantum case, this ambiguity in the
definition of $F$ is just a matter of normalization of the eigenfunction.

One can also consider the quantum counterpart of $T(p,q|g)$ which is a
Hamiltonian that gives rise to a Schr\"odinger equation w.r.t. the coupling
constant

\be
{\partial\psi\over\partial g}=i\hat T\psi
\ee
Therefore, the wavefunction $\psi$ can be also realized as a path integral
over the phase space variables $P(g),Q(g)$.

The generating functions $S(Q,q)$, $F(P,q)$ and $T(p,q)$ satisfy the quasiclassical
versions of these relations (when multiple derivatives of $S$, $F$ and $T$ are
neglected).
Exact definition of quantum evolution operators, including effects of
discrete spectra and identification of the spectra at different values of
coupling constants, i.e. precise definition of the spectral parameter
$\lambda$ in such a way that eq.(\ref{evo}) is diagonal in $\lambda$,
will be considered elsewhere.

\section{Some conceptual questions for further examination}

The previous section explained possible needs to study integrable systems
from the non-conventional perspective: in terms of the auxiliary
Hamiltonians $T(\vec p,\vec q|g)$, describing the associated Hamiltonian
flows in the space of coupling constants.
Some immediate problems were also mentioned, like

(i) criteria, distinguishing the Hamiltonians $T$ associated with
conventional integrable systems at the entire space of such Hamiltonians;

(ii) structure of algebra, generated by all the Hamiltonians $H_i$ and
$T_a$,
in which the $H_i$'s generate a commutative (Cartan) subalgebra;

(iii) duality relation between $\widetilde T$ and $T$;

(iv) relation between integrable systems and flat connections over the space
of coupling constants encoded in (\ref{PC}).

One can also add to this list:

(v) identification of the flows and Hamiltonians $T_a$ in terms of
Seiberg-Witten theory, \cite{SW,tauparexa};

(vi) treatment of $T$'s in terms of the theta-function formalism of
refs.\cite{dual,dual2};

(vii) the ``gauge-equivalence" of the Hamiltonians $T_a$,
corresponding to different choices of the constant-$\vec p,\vec q$
sections over the moduli space (i.e. with different quantities kept
constant when the couplings $g_a$ are varied);

(viii) the role of the distinguished ``flat-modulus" $\vec a$ \cite{SW},
for which the WDVV equations \cite{WDVV} hold;

(ix) further development of the notion of generalized $\tau$-function,
making use of the evolution operator $e^S$, introduced in the previous
section (see also the concluding section \ref{conc} below).

Since the subject is relatively new, we do not think it is time now to get into
further discussions of these (and many other) problems.
Instead in the remaining part of this paper we consider some simple
examples:
the Hamiltonians $T(\vec p,\vec q; g)$, associated with the flows
in the coupling constant space for some well-known integrable systems
like harmonic oscillator and
Calogero-Ruijsenaars-Double-elliptic (DELL)\footnote{
In fact, the treatment of the DELL case is very sensitive to the choices
mentioned in (vii) above, and we postpone serious discussion of the DELL case
to another occasion, in order not to draw attention away from the simple ideas
of the present paper.
}
family.

\section{Particular examples of Hamiltonian flows \label{exa}}

\subsection{Harmonic oscillator}

Perhaps surprisingly, even for this simple model one obtains rather
sophisticated formulas.

For harmonic oscillator $H(p,q|\omega) = \frac{1}{2}(p^2 + \omega^2q^2)$.
Let the frequency $\omega$ play the role of the coupling constant so that
$H^{(0)}(p) = \frac{1}{2}p^2$. Then

\be
P = \sqrt{p^2 + \omega^2q^2}, \\
Q = \frac{\sqrt{p^2 + \omega^2q^2}}{\omega}\arctan \frac{\omega q}{p}
\ee
Note that in this case the $\omega$-evolution possesses ``the conservation
law"

\be
\tan \frac{\omega Q}{P} = \frac{\omega q}{p}
\label{clho}
\ee

The inverse transformation looks like

\be
p = P\cos \frac{\omega Q}{P}, \\
q = \frac{P}{\omega} \sin \frac{\omega Q}{P}
\ee
i.e. the Hamiltonian of the dual flow is

\be
\widetilde H(\widetilde p,\widetilde q|\omega) =
\frac{\widetilde q}{\omega}\sin \frac{\omega \widetilde p}{\widetilde q}
\ee
(Note that with this definition $\widetilde H^{(0)}(p) = p$ and also note that
$p,q$-duality does not respect conventional dimensions of $p$ and $q$ so
that it should not be a surprise that $\omega q/p$ in $H$ got substituted
by $\omega \widetilde p/\widetilde q$ in $\widetilde H$.
Of course, $\widetilde p$ and $\widetilde q$ are nothing but $Q_\omega$ and $P_\omega$.)

The generator of $\omega$-evolution such that

\be
\left.\frac{\partial P}{\partial\omega}\right|_{p,q=const} =
\frac{\omega q^2}{\sqrt{p^2+\omega^2q^2}} =
\frac{P}{\omega}\sin^2\frac{\omega Q}{P} =
-\frac{\partial T}{\partial Q},
\nn\\
\left.\frac{\partial P}{\partial\omega}\right|_{p,q=const} =
\frac{p^2}{\omega^2\sqrt{p^2 + \omega^2q^2}}
\left(\frac{\omega q}{p} - \arctan\frac{\omega q}{p}\right)
= \nn\\=
\frac{P}{\omega^2}\cos^2\frac{\omega Q}{P}
\left(\tan\frac{\omega Q}{P} - \frac{\omega Q}{P}\right) =
\frac{\partial T}{\partial P},
\ee
is

\be
T = \frac{P^2}{4\omega^2}\sin\frac{2\omega Q}{P} -
\frac{PQ}{2\omega} = \frac{pq - PQ}{2\omega}
\label{Tho}
\ee
For the dual system we similarly have

\be
\left.\frac{\partial \widetilde P}{\partial\omega}\right|_{\widetilde p,\widetilde q=const} =
\frac{\widetilde q}{\omega^2}\left(\frac{\omega \widetilde p}{\widetilde q}
\cos \frac{\omega \widetilde p}{\widetilde q} - \sin\frac{\omega \widetilde p}{\widetilde q}\right)
= \nn \\ =
\frac{\widetilde Q}{\omega^2}\left(\frac{\omega\widetilde P}{\widetilde Q} -
\arctan\frac{\omega\widetilde P}{\widetilde Q}\right) =
-\frac{\widetilde T}{\widetilde Q}
\ee
and

\be
\widetilde T = \frac{-\omega\widetilde P\widetilde Q +
(\widetilde Q^2 + \omega^2\widetilde P^2)\arctan\frac{\omega \widetilde P}{\widetilde Q}
}{2\omega^2}
= \nn \\ =
\frac{\widetilde p\widetilde q - \widetilde P\widetilde Q}{2\omega} =
\frac{PQ -
pq}{2\omega} = -T
\ee

In order to obtain the generating function $S(Q,q|\omega)$ one needs first
to reexpress $p$ and $P$ through $q$ and $Q$. Unfortunately, the formulas are
transcendental. Instead, it is easy to express $p$ and $Q$ through
$q$ and $P$, this involves only elementary functions

\be
p = \sqrt{P^2 - \omega^2q^2}, \\ \nn
Q = \frac{P}{\omega}\arctan \frac{\omega q}{\sqrt{P^2 - \omega^2q^2}}
\ee
The corresponding generating function

\be
F(P,q|\omega) = \int pdq = \int\sqrt{P^2 - \omega^2q^2} dq
= \nn \\ =
\frac{1}{2}q\sqrt{P^2 - \omega^2q^2} +
\frac{P^2}{2i\omega} \log\left(i\frac{\omega q}{P} + \sqrt{1-\frac{\omega^2
q^2}{P^2}}\right)
\ee
so that the relation (\ref{efeF}) in quasiclassical approximation is

\be
e^{iF(\lambda, q)} =
\left(i\frac{\omega q}{\lambda} - \sqrt{1-
\frac{\omega^2q^2}{\lambda^2}}\right)^{\lambda^2/2\omega}
e^{{iq\over 2}\sqrt{\lambda^2 - \omega^2q^2}}
\sim \nn \\ \sim
e^{-\omega q^2/2} He_\nu(\sqrt{\omega}q)
\label{efeFho}
\ee
where
$\frac{\lambda^2}{2\omega} = \nu + \frac{1}{2}$
and the Hermite polynomials, satisfying

\be
(-\partial^2_x + x^2)e^{-x^2/2}He_\nu(x) = (2\nu+1)e^{-x^2/2} He_\nu(x)
\ee
are given by inverse Laplace transform of the generating function $e^{-x^2/4 +
t/x}$

\be
He_\nu(x) \sim \int \frac{dt}{t^{\nu + 1}}e^{-t^2/4 + tx}
\ee
Taking this integral in the saddle point approximation, one gets

\be
He_\nu(x) \sim \left(x-\sqrt{x^2-2(\nu+1)}\right)^{\nu+1}
e^{x^2/2+x/2\sqrt{x^2-2(\nu+1)}}
\ee
which, at large $\nu$, is in accordance with eq.(\ref{efeFho}).

\subsection{Generic Schr\"odinger-like systems}

For a single particle any dynamics is integrable,
and any Hamiltonian is canonically equivalent to the free one.
Therefore, it may make sense to look at the corollaries of our formalism
in such context. This can help to see the difference between models
which have and have not simple integrable generalizations to
the multiparticle case.

For $H = \sqrt{p^2 + g^2V(q)}$ the Hamiltonian $T$ can be found in the form
of a perturbation expansion:

\be\label{expan1}
T(p,q| g) = g\frac{T_0(q)}{p} + g^3\frac{T_1(q)}{p^3} + O(g^5)
\ee
where

\be
\frac{\partial T_0(q)}{\partial q} = -V(q), \ \ \
\frac{\partial T_1(q)}{\partial q} = \frac{1}{2}T_0\frac{\partial^2
T_0}{\partial q^2}
\label{expan}
\ee
There is no way to integrate explicitly these equations in the
general case.
However, for harmonic oscillator exact answer exists for the entire expansion,
eq.(\ref{Tho}). Similarly, exact answers are available
for potentials of the rational and trigonometric Calogero models (see below).
However, for elliptic Calogero model
with $V(q) \sim sn^{-2}(q|k)$ eq.(\ref{expan}) implies that
the generator $T(p,q|g)$ involves the
integral elliptic tangent (i.e. the Weierstrass zeta-function),
$Tn (q|k) = \int \frac{d\xi}{sn^2(\xi|k)} \sim \zeta(\hat \xi|\tau)$.\footnote{
However, the conservation laws of the $g$-flow involve the ordinary (algebraic) elliptic
tangents, $tn(\xi|k) = sn(\xi|k)/cn(\xi|k)$: say, in the $SU(2)$
elliptic-rational model (the dual of the elliptic Calogero system) \cite{dual}
$$
cn(P|k) = \alpha(q) cn(\beta(q)p|\gamma(q)),
$$
$$
q = \sqrt{Q^2 - \frac{g^2}{sn^2(P|k)}};
$$
$$
\alpha^2(q) = 1 - \frac{g^2}{q^2}, \ \
\beta^2(q) = k'^2 + k^2\alpha^2(q), \ \
\gamma(q) = \frac{k\alpha(q)}{\beta(q)}, \ \ k'^2 = 1 - k^2,
$$
the counterpart of (\ref{clho}) is
$$
Q\ tn(P|k) = q\ tn(\beta p|\gamma).
$$
The difference between the algebraic and integral tangents is that
$\frac{\partial}{\partial \xi)} tn(\xi|k) = \frac{dn(\xi|k)}{cn^2(\xi|k)}$
while $\frac{\partial}{\partial \xi)} Tn(\xi|k) = \frac{1}{cn^2(\xi|k)}$
}
Higher orders of the $g$ expansion contain multiple integrals of $Tn(\xi)$.
This reflects a more sophisticated group structure of elliptic models.

The simplest way to derive formulas (\ref{expan1})-(\ref{expan}) is to
construct the Hamiltonian $q=\widetilde H(P,Q)$ dual to
$\sqrt{p^2+g^2V(q)}$. The canonicity condition then gives

\be
{\partial \widetilde H(P,Q)\over\partial Q}={p(P,Q)\over P}
\ee
Therefore, the dual Hamiltonian can be obtained via solving the equation

\be\label{q}
Q=\int^q {d\xi\over\sqrt{1-g^2V(\xi)/P^2}}
\ee
w.r.t. $q$. Now taking the derivative of $P=\sqrt{p^2+g^2V(q)}$ w.r.t. $g$,
one obtains

\be\label{dT}
-{\partial T\over\partial Q}={g\over P}V(q(P,Q))
\ee
where $q(P,Q)$ is given by (\ref{q}). One easily obtains the expansion of these
two formulas in powers of $g$, in particular, reproducing (\ref{expan1})-(\ref{expan}).

\subsection{$SU(2)$ rational and trigonometric Calogero-Ruijsenaars models}

\subsubsection{The rational-rational case (rational Calogero model)}

In this case the single Hamiltonian is
$H = \frac{1}{2}\left(p^2 + \frac{g^2}{q^2}\right)$, thus $H^0(p) = \frac{1}{2}p^2$,
and

\be
P^2 = p^2 - \frac{g^2}{q^2}, \nn \\
q^2 = Q^2 - \frac{g^2}{P^2}
\ee
The rational Calogero model is self-dual.

One can express the dressed variables $P = P_g, Q = Q_g$ through the bare
ones, $p,q$

\be
P = \frac{1}{q}\sqrt{p^2q^2 - g^2}, \\
Q = \frac{pq^2}{\sqrt{p^2q^2 - g^2}}
\ee
The inverse transform is

\be
p = \frac{P^2Q}{\sqrt{P^2Q^2 - g^2}}, \\
q = \frac{1}{P}\sqrt{P^2Q^2 - g^2}
\ee
Note that, similarly to the case of harmonic oscillator, the
$g$-flow possesses a conservation law

\be
PQ = pq
\ee

The flow is generated by the Hamiltonian

\be
T(p,q;g) = \frac{1}{2}\log \frac{pq - g}{pq + g} = \frac{1}{2}\log \frac{PQ - g}{PQ + g}
\ee
This is obviously a symmetric function of $p$ and $q$, thus,
$T_D(p,q) = T(p,q)$ as it should be for the self-dual system.

In order to obtain the evolution operator $e^{iS}$, one should rewrite $T$
in terms of the $q,Q$-variables. The $p,P$-variables are then given by

\be
P^2 = \frac{g^2}{Q^2 - q^2}, \nn \\
p^2 = \frac{g^2Q^2}{q^2(Q^2-q^2)}.
\ee
The symmetricity (integrability) condition (\ref{symder}) is true

\be
\frac{\partial}{\partial q}\frac{1}{\sqrt{Q^2-q^2}} +
\frac{\partial}{\partial Q}\frac{Q}{q\sqrt{Q^2-q^2}} = 0,
\ee
and one obtains

\be
S = g\arccosh\frac{Q}{q} = g\log\left(\frac{Q - \sqrt{Q^2-q^2}}{q}\right) =
\frac{g}{2}\log \frac{Q - \sqrt{Q^2-q^2}}{Q + \sqrt{Q^2-q^2}}, \nn \\
\frac{\partial S(Q,q|g)}{\partial g} = T(p(Q,q),q|g)
\ee
In this case, $S$ is a simple linear function of $g$, and
$\partial S/\partial g = S/g$.

Similarly,

\be
F(P,q)=\sqrt{P^2q^2+g^2}+{g\over 2}\log {\sqrt{P^2q^2+g^2}-g\over
\sqrt{P^2q^2+g^2}+g},\\
\frac{\partial F(P,q|g)}{\partial g} = T(p(P,q),q|g)
\ee

One can also consider the transformation from the system given at
$(g_1,Q_1,P_1)$ to that given at $(g_2,Q_2,P_2)$. The corresponding
generating function

\be
S(Q_1,Q_2,g_1,g_2)=g_1\log
{g_1\sqrt{Q_1^2-Q_2^2}+\sqrt{g_1^2Q_1^2-g_2^2Q_2^2}\over Q_2} +
g_2\log
{g_2\sqrt{Q_2^2-Q_1^2}+\sqrt{g_2^2Q_2^2-g_1^2Q_1^2}\over Q_1}
\ee
is the function of both $g_1$ and $g_2$, while

\be
T={\partial S\over\partial g_2}={1\over 2}\log {P_2Q_2-g_2\over P_2Q_2+g_2}
\ee
is the function of only $g_2$.

In accordance with (\ref{efeS})

\be\label{qBes}
\psi_\lambda(q) \sim \int e^{iS(q,Q|g)} e^{iQ\lambda} dQ \sim qJ_{ig}(i\lambda q)
\ee
where $J_g(x)$ is the Bessel function. The exact quantum wavefunction, solving the
equation $(-\partial^2_x + \frac{g^2}{x^2})  \psi_\lambda(x)=
\lambda^2\psi_\lambda(x)$, is $\sqrt{x}J_\nu(i\lambda x)$, $\nu^2=-g^2+1/4$.
It coincides with (\ref{qBes}) in quasiclassical approximation.

For other members of the Calogero-Ruijsenaars family we just list some
relevant formulas, parallel to those in the rational-rational case.

\subsubsection{The rational-trigonometric case (trigonometric Calogero model)}

\be
P^2 = p^2 - \frac{g^2}{\sinh^2 q}, \\
\cosh^2 q =\cosh^2 Q\left(1 - \frac{g^2}{P^2}\right), \\
P\tanh Q = p\tanh q,\\
T = \frac{1}{2}\log \frac{P\tanh Q - g}{P\tanh Q + g}
= \frac{1}{2}\log \frac{p\tanh q - g}{p\tanh q + g}; \\
P^2 = \frac{g^2\cosh^2 Q}{\sinh^2Q - \sinh^2q}, \\
p^2 = \frac{g^2\sinh^2Q\cosh^2 q}{\sinh^2q(\sinh^2Q-\sinh^2q)};\\
\frac{\partial}{\partial q}\frac{\cosh Q}{\sqrt{\sinh^2Q-\sinh^2q}} +
\frac{\partial}{\partial Q}\frac{\sinh Q\cosh q}{\sinh q\sqrt{\sinh^2Q-\sinh^2q}} = 0
\ee
In this case $S$ is still a simple linear function of $g$

\be
S=g\arcsinh\left({\sinh Q\over\sinh q}\right)
\ee

\subsubsection{The trigonometric-rational case (rational Ruijsenaars model)}

\be
\cosh^2 P =\cosh^2 p\left(1 - \frac{\sinh^2\epsilon}{q^2}\right), \\
q^2 = Q^2 - \frac{\sinh^2\epsilon}{\sinh^2 P}, \\
Q\tanh P = q\tanh p,\\
T = \frac{1}{2}\log \frac{Q\tanh P - \tanh\epsilon}{Q\tanh P + \tanh\epsilon}
= \frac{1}{2}\log \frac{q\tanh p - \tanh\epsilon}{q\tanh p + \tanh\epsilon}; \\
\sinh^2 P = \frac{\sinh^2\epsilon}{Q^2 - q^2}, \\
\sinh^2 p = \frac{\sinh^2\epsilon Q^2}{(Q^2-q^2)(q^2-\sinh^2\epsilon)};\\
\frac{\partial}{\partial q}\arcsinh
\frac{g}{\sqrt{Q^2-q^2}} +
\frac{\partial}{\partial Q}\arcsinh \frac{gQ}
{\sqrt{(Q^2-q^2)(q^2 - \sinh^2\epsilon )}} = 0
\ee
In this case, $S$ is no longer a simple linear function of the coupling constant $\epsilon$.

\subsubsection{The trigonometric-trigonometric case (trigonometric Ruijsenaars model)}

\be
\cosh^2 P =\cosh^2 p\left(1 - \frac{\sinh^2\epsilon}{\sinh^2 q}\right), \\
\cosh^2 q =\cosh^2 Q\left(1 - \frac{\sinh^2\epsilon}{\sinh^2P}\right), \\
\tanh P\tanh Q = \tanh p\tanh q,\\
T = \frac{1}{2}\log \frac{\tanh P\tanh Q - \tanh\epsilon}{\tanh P\tanh Q + \tanh\epsilon}
= \frac{1}{2}\log \frac{\tanh p\tanh q - \tanh\epsilon}{\tanh p\tanh q + \tanh\epsilon}; \\
\sinh^2 P = \frac{\sinh^2\epsilon\cosh^2 Q}{\sinh^2Q - \sinh^2q}, \\
\sinh^2 p = \frac{\sinh^2\epsilon\sinh^2Q\cosh^2 q}{(\sinh^2Q-\sinh^2q)(\sinh^2 q - \sinh^2\epsilon)};\\
\frac{\partial}{\partial q}\arcsinh
\frac{\sinh\epsilon\cosh Q}{\sqrt{\sinh^2Q-\sinh^2q}} +
\frac{\partial}{\partial Q}\arcsinh\frac{\sinh\epsilon\sinh Q\cosh q}
{\sqrt{(\sinh^2Q-\sinh^2q)(\sinh^2 q - \sinh^2\epsilon)}} = 0
\ee

\subsection{$SU(N)$ rational Calogero model}

To give an illustration of what happens in the multiparticle case we
start here with the simplest case of the rational Calogero model.
More profound examples will be considered elsewhere.

The Hamiltonians $H_i(\vec p,\vec q) = Tr L^i$ are defined as traces of
powers of the Lax matrix $L_{ij} = p_i\delta_{ij} + g{1\over q_{ij}}$.
Accordingly $H_i^0(\vec p) = \sum_j p_j^i$.
The Hamiltonian flow within the family of rational Calogero models (along
the $g$ direction) is governed by

\be
T(\vec p,\vec q; g) = -g\sum_{ij}\frac{1}{p_{ij}q_{ij}} +O(g^3)
\ee

Technically, higher order terms can be done in the following way. As the first
step, one needs to find dual Hamiltonians. Note that dual Hamiltonians
technically each time emerge as necessary ingredient in constructing $T$.
Moreover, with these intermediate objects, constructing $T$ becomes
well-defined although often complicated procedure and reduces to
finding solutions of some partial differential equations. Indeed, first of
all the dual Hamiltonians emerge as solutions to partial differential canonicity
equations and then $T$ is obtained by integration the following set of
equations:

\be\label{rec}
\sum_i\left({\partial\widetilde H_k\over\partial Q_i}{\partial T\over\partial P_i}
-{\partial\widetilde H_k\over\partial Q_i}{\partial T\over\partial
P_i}\right)+{\partial\widetilde H_k\over \partial g}=0
\ee
These equations just follow from the fact that the dual Hamiltonians $\widetilde H_k$
are nothing but independent of the coupling constant $g$ coordinates $\{q_i\}$.

Now, having $\widetilde H_k$ calculated, one can easily construct a recursive
procedure of calculating $T$ as
series in $g$: $T=\sum g^{2k+1}T_k$. Indeed, note that, in the leading order in
$g$, $\widetilde H_k$ are just simple functions of $Q_i$, say, just $Q_k$
which do not depend on $g$. This means that plugging the power expansion of
$T$ into (\ref{rec}) gives a system of $N$ linear algebraic equations for
the derivatives $D_i^{(1)}\equiv\partial T_1/\partial P_i$. Solving these
equations, one should just integrate $D_i^{(1)}$, finally obtaining $T_1$.
Then one can use this calculated $T_1$ in order to find $T_2$ etc. Note that
each time there emerge non-trivial integrability conditions

\be
\partial D_i/\partial P_j=
\partial D_j/\partial P_i
\ee

Thus, the most difficult part of the calculation is to find dual Hamiltonians.
In the rational Calogero system, when the system is self-dual, the
calculation is very immediate. Similarly, the calculation can be easily done
for any member of the Calogero-Ruijsenaars-Dell family.

\section{Conclusion \label{conc}}

A peculiar feature of integrable systems is that interaction is introduced
by a canonical transformation of a free system.
In other words, the corresponding flows in the space of coupling
constants are Hamiltonian ones.
This observation should provide new insights into the structure
of exact (non-perturbative) partition functions, which are always
(generalized) $\tau$-functions of some integrable hierarchies
\cite{tauparrev,tauparexa}.

Let us remind that the partition function, arising after functional integration
over fields,

\be
\int D\varphi \exp \left({\cal A}(\varphi_0 + \varphi) +
i\sum_{\{k\}} t_{\{k\}} V_{\{k\}}(\varphi_0 + \varphi)\right) = \nn \\ =
\tau(t_{\{k\}}|\varphi_0) =
\ \langle \hat G(\varphi_0) P\exp\left(
i\sum_{\{k\}} \int \hat H_{\{k\}} dt_{\{k\}}\right)\rangle\ =
\ \langle \hat G e^{i\hat S} \rangle
\ee
is essentially unsensitive to the choice of original action
${\cal A}(\varphi)$, and the equivalence classes (of actions) can
be represented by integrable systems, i.e. by the systems where the
partition function is a matrix element of a group element $\hat G$ of
some group \cite{Hirota}. Our evolution operator in the space of coupling
constants,
$e^{i\hat S} =  P\exp\left(i\sum_{\{k\}} \int \hat H_{\{k\}} dt_{\{k\}}\right)$,
which performs a canonical transformation of integrable system into the
free one, as discussed in s.\ref{Haf} should also be a group element.
The moduli $\varphi_0$, parameterizing the boundary conditions and
vacuum states of the original theory are related to the (eigen)values
of the Hamiltonians $H_{\{k\}}$ (i.e., in our simple examples,
the dressed momenta $\vec P$).
The couplings (perturbations) $t_{\{k\}}$ (i.e. our $g$)
become time-variables of the $\tau$-function.
The fact that the space of relevant Hamiltonians
(the moduli space) is an orbit of the canonical
transformation group,
which can be alternatively considered as changing the couplings
can further clarifies the relation between
the times- and moduli-dependencies of the partition functions \cite{Kh}
and bring us closer to the long-awaited group-theoretical treatment
of {\it entire} $\tau$-function, its time-dependencies being associated with
the commutative (Cartanian) part of the algebra (generated by the
Hamiltonians $H_i$), while moduli-dependencies -- with its non-Cartanian
part (generated by $T_a$ and their commutators among themselves and with
$H_i$). See refs.\cite{group} for related approaches to this problem.

\section{Acknowledgements}

We acknowledge the hospitality of the University of British Columbia, where
the work was completed, and the support of the Nato Collaborative Linkage Grant SA
(PST.CLG.977361)5941. Our work is also supported by grants:
CRDF \#6531, INTAS 00-561
and also by the RFBR-01-02-17682a (A.Mir.) and RFBR-01-02-17488 (A.Mor.),
the Grant of Support for the Scientific Schools 96-15-96798 (A.Mir.)
and the Russian President's Grant 00-15-99296 (A.Mor.).

\end{document}